\documentclass[epj]{svjour}
\usepackage[numbers]{natbib}
\usepackage{amsmath}
\usepackage{graphicx}
\newcommand{\Vect}[1]{\boldsymbol{\rm #1}}
\newcommand{\Tens}[1]{\boldsymbol{\rm #1}}
\newcommand{\Vv}{\Vect{v}}
\newcommand{\Vvinf}{\Vect{v}^{\infty}}
\newcommand{\Vr}{\Vect{r}}
\newcommand{\Vt}{\Vect{t}}
\newcommand{\Ve}{\Vect{e}}
\newcommand{\Vn}{\Vect{n}}
\newcommand{\Vf}{\Vect{f}}
\newcommand{\Vfel}{\Vf^{\text{el}}}

\newcommand{\etain}{\eta_{\text{in}}}
\newcommand{\etaout}{\eta_{\text{out}}}

\newcommand{\Dd}{\textrm{d}}
\newcommand{\Jump}[1]{\left[#1\right]^{\text{out}}_{\text{in}}}
\newcommand{\Vnabla}{\Vect{\nabla}}
\DeclareMathOperator{\sign}{sign}

\DeclareMathOperator{\Imaginary}{Im}

\begin{document}

\title{Two-Dimensional Fluctuating Vesicles in Linear Shear Flow}

\author{Reimar Finken\inst{1}\thanks{\email{finken@theo2.physik.uni-stuttgart.de}} 
\and Antonio Lamura\inst{2}\thanks{\email{a.lamura@ba.iac.cnr.it}} 
\and Udo Seifert\inst{1}\thanks{\email{useifert@theo2.physik.uni-stuttgart.de}} 
\and Gerhard Gompper\inst{3}\thanks{\email{g.gompper@fz-juelich.de}}}

\institute{%
\inst{1} 2. Institut f{\"u}r Theoretische Physik, Pfaffenwaldring  57/III, 
Universit{\"a}t Stuttgart, 70550 Stuttgart, Germany \\ 
\inst{2} Istituto Applicazioni Calcolo, Consiglio Nazionale delle Ricerche (CNR),
Via Amendola 122/D, 70126 Bari, Italy \\ 
\inst{3} Forschungszentrum J{\"u}lich GmbH, 
Institut f{\"u}r Festk{\"o}rperforschung,
52425 J{\"u}lich, Germany
}

\date{\today}

\abstract{
  The stochastic motion of a two-dimensional vesicle in linear shear flow is
  studied at finite temperature. In the limit of small deformations from a
  circle, Langevin-type equations of motion are derived, which are highly
  nonlinear due to the constraint of constant perimeter length. These
  equations are solved in the low temperature limit and using a mean field
  approach, in which the length constraint is satisfied only on average. The
  constraint imposes non-trivial correlations between the lowest deformation
  modes at low temperature. We also simulate a vesicle in a hydrodynamic
  solvent by using the multi-particle collision dynamics technique, both
  in the quasi-circular regime and for larger deformations, and compare the
  stationary deformation correlation functions and the time autocorrelation
  functions with theoretical predictions. Good agreement between theory and
  simulations is obtained.
}



\maketitle


\section{Introduction}
\label{sec:introduction}
The dynamics of soft objects such as drops, capsules and cells in flow
represents a long-standing problem in science and engineering, but has
received increasing interest recently, in particular due to its relevance to
biological, medicinal and microfluidic applications. This problem is
challenging from a theoretical point of view, because the shape of these
objects is not given \emph{a priori}, but determined dynamically from a
balance of interfacial forces with fluid stresses. Improved experimental
methods have revealed intriguing new dynamical shape transitions due to the
presence of shear flow. The phenomenology of the dynamical behavior depends
distinctively on the specific soft object immersed in the flow with fluid
bilayer vesicles and elastic microcapsules as the most prominent classes.

Fluid bilayer vesicles assume a stationary tank-treading shape in linear shear
flow, if there is no viscosity contrast between interior and exterior fluid
\cite{kraus1996}. If the interior fluid or the membrane becomes more viscous,
a transition to a tumbling state can occur
\cite{biben2003a,beaucourt2004b,rioual2004a,noguchi2004,noguchi2005a,vlahovska2007}.
Tank-treading was observed experimentally in infinite shear flow
\cite{haas1997,kantsler2005a} and for vesicles interacting with a rigid wall
\cite{lorz2000,abka02}, where a dynamic lift occurs
\cite{seifert1999b,cantat1999a,sukumaran2001,beaucourt2004a}. The
tank-treading to tumbling transition was observed for the first time
convincingly in an experiment only very recently \cite{kantsler2006a}. In addition to the
tank-treading to tumbling transition, an oscillating motion was predicted
theoretically \cite{misbah2006} and observed experimentally
\cite{kantsler2006a} and in simulations \cite{noguchi2007}. This type of
motion has alternatively been called vacillating-breathing \cite{misbah2006},
swinging \cite{noguchi2007}, or trembling \cite{kantsler2006a,lebedev2007}.
The theoretical description has been extended recently beyond first order in
the shear rate \cite{noguchi2007,lebedev2007,danker2007}.

At finite temperature, stochastic fluctuations of the membrane due to thermal
motion affect the motion of the object. Due to the dissipative nature of the
hydrodynamic interactions, vesicles in shear flow form a non-trivial
model system for studying non-equilibrium stochastic dynamics. Since the
effect of thermal noise on the transitions between the different modes of
motion in general is a challenging task, in this paper we concentrate on the
stochastic motion in the stationary tank-treading state. Our theoretical
approach is similar to that of Ref. \cite{seifert1999a}, where stochastic equations of
motion were derived for quasi-spherical vesicles.

Most numerical methods solving the equation of motion of vesicles or capsules
\cite{kraus1996,biben2003a} operate in the absence of thermal forces. An
exception, which naturally includes thermal noise, is multi-particle collision
dynamics (MPC), also known as stochastic rotation dynamics (SRD)
\cite{malevanets1999,malevanets2000,kikuchi2003,yeomans2006}. In this method,
the fluid part is modeled on a particle rather than a continuum level. The
microscopic equations of motion for the effective fluid are chosen to be
evaluated efficiently on the one hand, and on the other hand to lead to the
correct macroscopic hydrodynamics. This method has successfully been applied
to flow around rigid objects \cite{lamura2001a,lamura2002}, polymers
\cite{kikuchi2005,ripoll2006} and viscous vesicles
\cite{noguchi2004,noguchi2005a,noguchi2005,noguchi2007}. We employ the MPC
simulation method to compare our theoretical predictions of correlation
functions, inclination angles, and tank-treading frequencies with simulation
data of vesicles. In order to obtain good statistics, we focus here on
two-dimensional (2d) vesicles with a linear boundary.

The paper is organized as follows: After formulating the problem in section \ref{sec:problem}, we develop
nonlinear stochastic equations of motion for quasi-circular vesicles in
section \ref{sec:quasi-circ-appr}. These are solved approximately using a mean field approach and a low
temperature expansion in section \ref{sec:appr-solut}. We also present the 2d version of the deterministic
Keller-Skalak theory \cite{keller1982} in section \ref{sec:keller-skalak-theory}, which takes into account the influence of the vesicle
shape on the flow. The simulation method used is discussed in section
\ref{sec:simulation-method}. Finally we compare the calculations with simulation
data in section \ref{sec:comp-with-simul} and discuss our results.

\section{Problem formulation}
\label{sec:problem}

We consider a model 2d vesicle immersed in a fluid of viscosity $\etaout$ with
a 1d membrane boundary surrounding a fluid of viscosity $\etain$ and at finite
temperature $T$. Due to the incompressibility of the membrane and of the enclosed
fluid, the area $A_{0}$ and the length $L_{0}$ of the membrane are constants.
The membrane resists deformation with a bending rigidity $\kappa$, which is
defined rigorously below in section \ref{sec:const-equat-membr}. The fixed
area defines a length scale
\begin{equation}
  \label{eq:1}
  R_{0} \equiv \sqrt{\frac{A_{0}}{\pi}},
\end{equation}
which can be used to define a number of dimensionless quantities. In the
following, we use the excess length
\begin{equation}
  \label{eq:2}
  \Delta \equiv \frac{L_{0}}{R_{0}} - 2\pi
\end{equation}
and the dimensionless viscosity contrast
\begin{equation}
  \label{eq:3}
  \lambda \equiv \frac{\etain}{\etaout}.
\end{equation}
Alternatively one can derive a length $R^{*}\equiv L_{0}/(2\pi)$ from the length
constraint, and use it to define a reduced area $A^{*}\equiv A_{0}/(\pi
{R^{*}}^{2})$. The reduced area is connected to the excess length by 
\begin{equation}
  \label{eq:4}
  A^{*}= \left(1+\frac{\Delta}{2\pi}\right)^{-2}.
\end{equation}

In a
quiescent fluid, thermal stochastic forces acting on the membrane lead to a
fluctuating shape, where the probability of any specific deformation can be
calculated using the Boltzmann weight corresponding to the deformation energy
$\mathcal{H}[\Vr]$. If an external flow field $\Vvinf$ is switched on, the system
ceases to be in equilibrium, and the statistical weight of a deformation
cannot be calculated \emph{a priory} using Boltzmann weights. We first
derive the force balance governing the motion of a vesicle in stochastic
Stokes flow, before we simplify the equations of motion in the limit of small
deformations from a circular shape.

\subsection{Constitutive equation of the membrane}
\label{sec:const-equat-membr}
We employ conventions of differential geometry following
Ref.~\cite{carmo1976}. The shape of the vesicle is given by the shape function
$\Vr(s)$, where $0\le s \le L$ denotes the arc length. The tangent vector
$\Vt(s)\equiv \Dd \Vr(s)/\Dd s$ is of unit length. The unit normal vector $\Vn(s)$ is
defined to point to the outside of the vesicle, and the orientation is chosen
such that the pair $(\Vn,\Vt)$ forms a right handed system. The curvature $k(s)$
is defined via the relation
\begin{equation}
  \Dd \Vt(s) / \Dd s = - k(s) \Vn(s).\label{eq:curv}
\end{equation}
The 2d analog of the bending energy of a certain
membrane deformation is given by the Helfrich term \cite{helfrich1973}
\begin{equation}
  \mathcal{H}_{\kappa}[\Vr] \equiv \frac{\kappa}{2} \int_{0}^{L} \Dd s \,
  k(s)^{2},
\end{equation}
which corresponds also to the bending energy of a semi-flexible polymer
\cite{kratky1949}.  Note that for 3d vesicles, a spontaneous
curvature $C_0$ can appear in the bending energy for intrinsically asymmetric
monolayers or asymmetric liquid environments. In 2d vesicles, we can ignore
the spontaneous curvature, since it shifts the bending energy only by a
topological constant, much like the Gaussian curvature contribution to the
curvature energy can be ignored in 3d.

All deformations of the vesicle must preserve the length $L$. In addition, the
fluid membrane is locally incompressible. This is ensured by introducing the
tension $\sigma(s)$ as a Lagrange parameter. The total energy thus reads
\begin{equation}
  \mathcal{H}[\Vr]= \mathcal{H}_{\kappa}[\Vr] + \int_{0}^{L} \Dd s\, \sigma(s).
\end{equation}
From the Euler-Lagrange equations we can deduce the force acting on the
membrane
\begin{equation}
  \label{eq:5}
  \Vfel = - \frac{\delta \mathcal{H}[\Vr]}{\delta \Vr} = \Vt\left(\sigma' + 2 \kappa k k'\right)+
  \Vn\left(\frac{\kappa}{2} k^{3}-\kappa k'' - k \sigma\right).
\end{equation}
Here the prime denotes a derivative with respect to the arc length $s$.

\subsection{Stochastic Stokes flow}
\label{sec:stoch-stok-flow}
The elastic forces given by Eq.~(\ref{eq:5}) are balanced by hydrodynamic
forces mediated by the surrounding fluid. The motion of the fluid and the
vesicle is only considered on time scales on which the fluid is 
incompressible, i.e. 
\begin{equation}
  \label{eq:incfluid}
  \Vnabla \cdot \Vv = 0. 
\end{equation}
The length and time scales in typical experiments and simulations is such that
the Reynolds number is very small. We only consider fluctuations on time
scales on which the inertial term in the Navier-Stokes equation can be
neglected. The velocity field $\Vv$ of the fluid is then subject to the
steady stochastic Stokes equation \cite{landau1959fluid}
\begin{equation}
  \label{eq:navier-stochastic}
   - \Vnabla p + \eta_\alpha \Delta \Vv + \Vnabla \cdot \Tens{s} = 0,
\end{equation}
where the thermal stress tensor $\Tens{s}(\Vect{x})$ is assumed to be a
Gaussian random variable with zero mean and correlations
\begin{equation}
  \begin{split}
    \langle s_{ik} \rangle &= 0\\
    \langle s_{ik}(\Vect{x}_{1},t_{1}) s_{lm}(\Vect{x}_{2},t_{2})\rangle &= 2 k_{B} T
    \eta_{\alpha} \delta(\Vect{x}_{1}-\Vect{x}_{2}) \delta(t_{1}-t_{2})\\
    &\quad \times [\delta_{il} \delta_{km} + \delta_{im} \delta_{kl}].
  \end{split}
\end{equation}
Here $\alpha\in \{\text{in},\text{out}\}$  indicates the inner or outer fluid. Instead
of calculating the stochastic velocity field of the flow, we only calculate the
deterministic part of $\Vv$:
\begin{equation}
  \label{eq:navier}
   - \Vnabla p + \eta_\alpha \Delta \Vv = 0.
\end{equation}
At the vesicle membrane we must have force balance between the deterministic
and stochastic part of the hydrodynamic force and the elastic forces
\begin{equation}
  \Vf^{\text{el}} + \Jump{\Tens{T}\cdot\Vn} + \Tens{s}\cdot \Vn = 0.\label{eq:6}
\end{equation}
Here, $\Tens{T}$ denotes the deterministic hydrodynamic stress tensor with
Cartesian components
\begin{equation}
  \label{eq:7}
  T_{ik} \equiv -p \delta_{ik} + \eta_{\alpha}[\partial_{i} v_{k} + \partial_{k}v_{i}].
\end{equation}
Far away from the vesicle the velocity field assumes the externally given
values
\begin{equation}
  \label{eq:8}
  \Vv(\Vect{x}) \rightarrow \Vv^{\infty}(\Vect{x}),\quad |\Vect{x}| \rightarrow \infty,
\end{equation}
which is ensured by separating an induced part from the velocity field
\begin{equation}
  \label{eq:9}
  \Vv \equiv \Vv^{\infty}+\Vv^{\text{ind}},
\end{equation}
and requiring that the induced part drops to zero far away from the vesicle.
Assuming no-slip boundary conditions, the vesicle is advected by the flow,
which implies
\begin{equation}
  \label{eq:10}
  \partial_{t} \Vr(s,t) = \Vv(\Vr(s,t),t).
\end{equation}
Here the dynamics still depends implicitly on $\sigma(s)$, which has to be
chosen such that $s$ remains the arc length, ensuring incompressibility. 
Eqs.~(\ref{eq:5},\ref{eq:navier}--\ref{eq:10}) determine the stochastic
motion of the vesicle.

\section{Quasi-circular approximation}
\label{sec:quasi-circ-appr}

These equations can be simplified considerably if we restrict ourselves to
vesicle shapes close to the circle. We parameterize the shapes as a function
of the polar angle $\phi$
\begin{equation}
  \label{eq:11}
  \Vr(\phi) = R_{0} \Ve_{r}(\phi)  (1+u(\phi)),
\end{equation}
and consider small distortions $u$. The deformation amplitude $u(\phi)$
is a real periodic function of $\phi$ and can therefore be expanded into complex Fourier modes
\begin{equation}
  \label{eq:fourier}
  u(\phi) \equiv  \sum_{m = -\infty}^{\infty} u_m \frac{\exp(i m \phi)}{\sqrt{2\pi}}.
\end{equation}
For comparison with simulation data described below, an expansion into a
real Fourier series is advantageous. We therefore also employ the expansion
\begin{equation}
  \label{eq:12}
  u(\phi) \equiv a_{0} + \sum_{m=1}^{\infty} a_{m} \cos(m\phi) +
  \sum_{m=1}^{\infty} b_{m} \sin(m \phi).
\end{equation}
The real Fourier coefficients $a_{m}, b_{m}$ are connected with the complex
Fourier coefficients $u_{m}$ via ($m\neq 0$)
\begin{equation}
  \label{eq:13}
  u_{m} = \sqrt{\frac{\pi}{2}}\left(a_{m} - i b_{m} \right).
\end{equation}
Area conservation fixes $u_{0}$ in terms of the other  $u_{m}$ 
\begin{equation}
  u_{0} = - \frac{1}{\sqrt{8\pi}} \sum_{m \neq 0} |u_{m}|^{2}.
\end{equation}
This relation will be used throughout the paper, and from now on sums over $m$
exclude the $m=0$ term. The contour length $L$ of the membrane is calculated to second order
in $u$ to be
\begin{equation}
  L = 2 \pi R_{0} + \frac{R_{0}}{2} \sum_{m\neq 0} (m+1)(m-1) |u_{m}|^{2}.
\end{equation}
Hence the excess length $\Delta$ reads
\begin{equation}
  \Delta = \frac{1}{2} \sum_{m\neq 0} \left(m^2-1\right) |u_{m}|^{2}.\label{eq:length}
\end{equation}
Finally, the local curvature $k$ evaluates to
\begin{equation}
  \label{eq:14}
   R_{0} k(\phi) = 1 - u''(\phi) = 1 + \sum_{m\neq 0} m^{2} u_{m} \frac{\exp(i m\phi)}{\sqrt{2\pi}}.
\end{equation}
This leads to the bending energy (ignoring constant terms)
\begin{equation}
  \label{eq:15}
  \mathcal{H}_{\kappa} = \frac{\kappa}{4 R_{0}}\sum_{m\neq 0}
  \left(m^{2}-1\right)\left(m^{2}-3/2\right) |u_{m}|^{2}.
\end{equation}
We now add the global length constraint (\ref{eq:length}) with a Lagrangian
multiplier
\begin{equation}
  \label{eq:16}
  \Sigma \equiv \kappa \sigma / R_{0}^{2}
\end{equation}
to the quadratic part of the bending energy. This leads to a quadratic
expression for the total energy (\ref{eq:52})
\begin{equation}
  \label{eq:17}
  \mathcal{H} = \frac{\kappa}{2 R_{0}} \sum_{m \neq 0} E_{m}(\sigma) |u_{m}|^{2}.
\end{equation}
with
\begin{equation}
  \label{eq:18}
  E_{m}(\sigma) \equiv (m+1)(m-1)[m^{2}-3/2+ \sigma].
\end{equation}
The bending forces (\ref{eq:5}) are determined by the deformation amplitudes
$u_{m}$ and by the instantaneous tension
\begin{equation}
  \label{eq:19}
  \Sigma(\phi) \equiv \frac{\kappa}{R_{0}}\left(\sigma + \sum_{m\neq 0} \sigma_{m}
    \frac{\exp(i m\phi)}{\sqrt{2\pi}}\right).
\end{equation}
The homogeneous tension $\kappa \sigma/R_{0}$ has already been included into the
energy \eqref{eq:17}.

\subsection{Velocity field}
\label{sec:velocity-field}
In polar coordinates, the general solution of Stokes' equation can be
expanded into the fundamental modes \cite{happel1983}
\begin{equation}\label{eq:lambmodes}
  \begin{array}{rlrl}
    \Vv^{\Phi,\pm}_{m} & \equiv \Vnabla (r^{\pm |m|}\exp(i m
    \phi))/\sqrt{2\pi}, \\
    p^{\Phi,\pm}_{m} & \equiv 0,\\
    \Vv^{P,\pm}_{m} &\equiv \frac{1}{2(1 \pm
      |m|)\sqrt{2\pi}}\left(\frac{|m|\pm 2}{2|m|} r^{2} \Vnabla(r^{\pm |m|}
      \exp(i m \phi))\right.\\
    &\quad \left.- \Vr r^{\pm |m|} \exp(i m \phi)\right),\\
    p^{P,\pm}_{m} &\equiv \eta_\alpha r^{\pm |m|} \exp(i m \phi)/\sqrt{2\pi}.
  \end{array}
\end{equation}
In this representation the cases $|m|=1$ and $|m|=0$ are special and
have to be treated separately. They correspond to constant flow and
rotational flow, respectively. The deeper reason why these are special
cases is the Stokes paradox \cite{happel1983}.  It follows from the boundary conditions that the induced
velocity field on the inside must be composed of ``$+$'' modes, and of
``$-$'' modes on the outside.
 
The corresponding hydrodynamic stress tensor reads in $(r,\phi)$
components
\begin{equation}
  \begin{split}
    \Tens{T}^{\Phi,\pm}_{m} &=2 \eta r^{\pm|m|-2} \frac{\exp(i m \phi)}{\sqrt{2
        \pi}}\\
    &\quad \times
    \begin{pmatrix}
      |m|(|m| \mp 1) &  i m (\pm |m| - 1)\\
      i m (\pm |m|-1) & - |m|(|m| \mp 1)\\
    \end{pmatrix}
  \end{split}
\end{equation}
and
\begin{equation}
  \Tens{T}^{P,\pm}_{m} = \eta r^{\pm |m|} \frac{\exp(i m \phi)}{2\sqrt{2\pi}}
  \begin{pmatrix}
    \pm |m| -2 & i m\\
    i m & \mp |m| - 2\\
  \end{pmatrix}.
\end{equation}
We can now express the 2d Oseen tensor in spectral components. The radial and
polar components of the fluid velocity and hydrodynamic force at the reference
circle are expanded into Fourier modes analogous to the expansion
\eqref{eq:fourier}. The velocity field at the reference sphere together with
the boundary conditions uniquely
determines the expansion (\ref{eq:lambmodes}). From the spatial velocity field the hydrodynamic
force $\Vf^{\pm} \equiv \Tens{T}^{\pm} \cdot \Vn$ can be calculated, leading to
\begin{equation}
  \begin{pmatrix}
    f^{r,\text{ind}}_{m}\\ f^{\phi,\text{ind}}_{m}
  \end{pmatrix}=
  \frac{\eta_{\text{in}} + \eta_{\text{out}}}{R_{0}}
  \begin{pmatrix}
    2|m| & 2 i \sign(m)\\
    -2 i \sign(m) & 2 |m|
  \end{pmatrix}
  \cdot
  \begin{pmatrix}
    v^{r,\text{ind}}_{m}\\ v^{\phi,\text{ind}}_{m}
  \end{pmatrix}.
\end{equation}

\subsection{External flow}
\label{sec:external-flow}
In the absence of the vesicle the applied external flow must be regular
everywhere. Therefore apart from constant flow and constant rotation only the
``+'' modes contribute in the expansion (\ref{eq:lambmodes}). To avoid
 the intricacies of the Stokes paradox, we neglect the possibility
of constant flow. A general expansion of the external flow therefore reads 
\begin{equation}
  \label{eq:20}
  \Vv^{\infty} = \sum_{m} \Phi^{\infty}_{m} \Vv^{\Phi,+}_{m} + P_{m}^{\infty}
  \Vv^{P,+}_{m} + \Omega r \Ve_{\phi}.
\end{equation}
The last term in this expansion corresponds to rotational flow with the
vorticity $\Omega$. For a finite viscosity contrast there is a jump in the
traction
\begin{equation}
  \begin{split}
    \begin{pmatrix}
      f^{r,\infty}\\ f^{\phi,\infty}
    \end{pmatrix}&= \sum_{m} \frac{\exp (i m \phi)}{\sqrt{2\pi}}
    (\eta_{\text{in}} - \eta_{\text{out}})\\ &\quad \times\left[\begin{pmatrix}
        2|m|(|m|-1)\\
        2 i m (|m|-1)
      \end{pmatrix} \Phi^{\infty}_{m} +
      \begin{pmatrix}
        |m|/2-1\\
        i m/ 2
      \end{pmatrix}
      P^{\infty}_{m} \right]
  \end{split}
\end{equation}
For the specific case of external linear shear flow
\begin{equation}
  \begin{split}
    \Vv^{\infty} &= \dot{\gamma} y \Ve_{x} = (\dot{\gamma}/2)
    (y\Ve_{x}+x\Ve_{y}) - (\dot{\gamma}/2) (x \Ve_{y}-y\Ve_{x})\\ &=
    -(i\dot{\gamma}\sqrt{2\pi}/8)
    \left[\Vv^{\Phi,+}_{2}-\Vv^{\Phi,+}_{-2}\right] - (\dot{\gamma}/2) r
    \Ve_{\phi},
  \end{split}
\end{equation}
we can read off the only non-vanishing components
\begin{equation}
  \label{eq:21}
  \begin{split}
    \Phi^{\infty}_{2} &= - \Phi^{\infty}_{-2} = -i \sqrt{2\pi}\dot{\gamma} /8,\\
    \Omega &= -\dot{\gamma}/2.
  \end{split}
\end{equation}
We will also use the dimensionless shear rate 
\begin{equation}
  \label{eq:22}
  \chi \equiv \dot{\gamma} \frac{\etaout R_{0}^{3}}{\kappa}
\end{equation}
and vorticity
\begin{equation}
  \label{eq:23}
 \tilde{\Omega} \equiv \Omega  \frac{\etaout R_{0}^{3}}{\kappa} = -\frac{\chi}{2}.
\end{equation}

\subsection{Incompressibility condition}
\label{sec:incompr-cond}
The flow at the vesicle membrane is subject to the incompressibility condition
$D_{t} \sqrt{g} = 0$, which can be cast in the equivalent form $\Vt \cdot
\partial_{\phi} \Vv(\Vr(\phi)) = 0$. To leading order in the deformation, this
condition reads 
\begin{equation}
  \label{eq:inc}
  v_{r}(R_{0}) + \partial_{\phi} v_{\phi}(R_{0}) = 0.
\end{equation}
Separating the induced flow from the external flow, we have in Fourier
components
\begin{equation}
  \begin{split}
    v^{r,\text{ind}}_{m}(R_{0}) &+ i m v^{\phi,\text{ind}}_{m}(R_{0}) = -
    v^{r,\infty}_{m}(R_{0}) - i m v^{r,\infty}_{m}(R_{0})\\
    &=   |m|(|m|-1)R_{0}^{|m|-1}\Phi^{\infty}_{m} +
    \frac{m^{2}}{4(|m|+1)}P^{\infty}_{m}.
  \end{split}
\end{equation}
Using this relation, we can eliminate $v^{\phi}_{m}$ and obtain
\begin{equation}
  \begin{split}
    f^{r}_{m} &= 2 \frac{\eta_{\text{out}}}{R_{0}}(\lambda+1)
    \frac{m^{2}-1}{|m|} v^{r,\text{ind}}_{m} \\
    &\quad + \Phi^{\infty}_{m}2(|m|-1)
    \frac{\eta_{\text{out}}}{R_{0}} \left[|m|(\lambda - 1) +
      (\lambda+1)\right] + P^{\infty}_{m}.
  \end{split}
\end{equation}

\subsection{Equation of motion}
\label{sec:equation-motion}
Neglecting the thermal fluctuating forces for the moment, we can derive a
deterministic equation of motion. The force balance leads to 
\begin{equation}
  \label{eq:24}
  v^{\text{ind},r}_{m} = - (\kappa /\eta_{\text{out}}R_{0}^{2}) \Gamma_{m} E_{m}(\sigma) u_{m} +
  B_{m} \Phi^{\infty}_{m} R_{0}
\end{equation}
with
\begin{equation}
  \label{eq:25}
   \Gamma_{m} \equiv \frac{|m|}{2(\lambda+1)(m^{2}-1)}
\end{equation}
and
\begin{equation}
  \label{eq:26}
  B_{m} \equiv-   \frac{m^{2}(\lambda
    - 1) + |m|(\lambda+1)}{(\lambda+1)(|m|+1)}.
\end{equation}
From the induced velocity, we obtain the radial component of the full velocity
field
\begin{equation}
  \label{eq:27}
  v^{r}_{m} = v^{r,\text{ind}}_{m}+v^{r,\infty}_{m} = v^{r,\text{ind}}_{m} +
  |m| \Phi^{\infty}_{m}. 
\end{equation}
The advection equation then reads (cf. Ref.~\cite{seifert1999a})
\begin{equation}
  \label{eq:eomdet}
  \partial_{t} u_{m} =  i \Omega m u_{m} - (\kappa /\eta_{\text{out}} R_{0}^{3})
  \Gamma_{m} E_{m}(\sigma) u_{m} + D_{m}\Phi^{\infty}_{m}
\end{equation}
with
\begin{equation}
  \label{eq:28}
D_{m} \equiv \frac{2 m}{\lambda+1}
\end{equation}
At non-zero temperature, thermal forces must be taken into account in the force
balance. The deterministic equation of motion (\ref{eq:eomdet}) then becomes a
Langevin equation
\begin{equation}
\label{eq:langevin}
  \partial_{t} u_{m} =  i \Omega m u_{m} - (\kappa /\eta_{\text{out}}R_{0}^{3})
  \Gamma_{m} E_{m}(\sigma) u_{m} + D_{m}\Phi^{\infty}_{m} + \zeta_{m}.
\end{equation}
The form of the thermal noise $\zeta_{m}$ can be obtained directly from the
noise term in Eq.~\eqref{eq:6}. It is much easier, however, to determine $\zeta_{m}$
from the Einstein relation, which must be valid in equilibrium. We assume that
the equilibrium noise is valid also
for non-vanishing shear flow and choose
\begin{equation}
  \label{eq:29}
  \langle \zeta_{m}(t) \zeta_{m'}(t') \rangle = 2 (k_{B} T \Gamma_{m} /\etaout
  R_{0}^{3})\delta_{m,-m'} \delta(t-t').
\end{equation}
Eq.~(\ref{eq:langevin}) is the correct stochastic equation of motion
for the vesicle deformation modes $u_{m}$. The tension $\sigma$ is at each
instance determined such that the length constraint (\ref{eq:length}) is
fulfilled. Taking the time derivative of Eq.~(\ref{eq:length}) and
using Eq.~(\ref{eq:langevin}), we can solve for the tension
\begin{equation}
  \label{eq:tenscons}
  \begin{split}
    \sigma &= \left[\sum_{m\neq 0}
      (m^{2}-1)^{2}\Gamma_{m}|u_{m}|^{2}\right]^{-1}\sum_{m\neq
        0}(m^{2}-1) \Bigl[i \Omega m |u_{m}|^{2} \Bigr.\\
    &\quad \Bigl. - \frac{\kappa}{\eta_{\text{out}}R_{0}^{3}} \Gamma_{m} E_{m}(0) |u_{m}|^{2} + D_{m}
      u_{m}^{*}\Phi^{\infty}_{m} + u_{m}^{*}\zeta_{m}\Bigr].
  \end{split}
\end{equation}
When this expression is inserted back into Eq.~(\ref{eq:langevin}), the
resulting noise term becomes dependent on the instantaneous values of the
$u_{m}$. While such non-linear noise terms hold interesting physics, we first
concentrate on tractable approximate solutions to the stochastic equation of
motion. 

\section{Approximate solutions}
\label{sec:appr-solut}

\subsection{Mean-field treatment}
\label{sec:finite-temperature}
At finite temperature, higher-order modes are excited by stochastic thermal
forces and therefore cannot be neglected. The full non-linear set of Langevin
equations (\ref{eq:langevin}) in combination with the expression
(\ref{eq:tenscons}) for $\sigma$ is too complex to admit a general solution.
We can, however, gain further insight in the tank-treading regime using a
mean-field description. We replace the fluctuating tension $\sigma$ in
Eq.~(\ref{eq:langevin}) by a constant, which has to be determined
self-consistently from the length constraint. The Langevin equations
(\ref{eq:langevin}) then become linear and decouple. In the stationary state,
only the $m=2$ deformations have a finite mean,
\begin{equation}
  \langle u_{2} \rangle = \frac{\etaout R_{0}^{3}}{\kappa} \frac{D_{2}}{\Gamma_{2}E_{2}(\sigma) + i \chi} \Phi^{\infty}_{2}.\label{eq:umean}
\end{equation}
On average, the vesicle is elliptical. As a measure of the deformation from
the circle we define the Taylor deformation parameter
\begin{equation}
  \label{eq:taylor}
  D\equiv \frac{L-S}{L+S},
\end{equation}
where $L$ and $S$ denote the long and short axis of the ellipse. In the
mean-field treatment we have
\begin{equation}
  \label{eq:30}
  D = \frac{2}{3}
  \frac{\chi}{\left[(5/2+\sigma)^{2} +
      9 \chi^{2}(1+\lambda)^{2}\right]^{1/2}}.
\end{equation}
The inclination angle is obtained from Eq.~(\ref{eq:umean}) 
\begin{equation}
  \label{eq:31}
  \Theta = \frac{1}{2}\arctan
  \frac{5/2+\sigma}{\chi(1+\lambda)}.
\end{equation}
The deviations from the mean 
\begin{equation}
\delta u_{m} \equiv u_{m} - \langle
u_{m}\rangle\label{eq:32}
\end{equation}
obey the homogeneous Langevin equation
\begin{equation}
  \label{eq:33}
    \partial_{t} \delta u_{m} =  i m \Omega \delta u_{m} - (\kappa /\etaout R_{0}^{3})
  \Gamma_{m} E_{m}(\sigma) \delta u_{m} + \zeta_{m}.
\end{equation}
The stationary noise correlations are best evaluated using a time Fourier
transform 
\begin{equation}
\widehat{\delta u}(\omega) \equiv \int dt \exp(- i \omega t)
\delta u(t)\label{eq:34},
\end{equation}
leading to
\begin{equation}
  i \omega \widehat{\delta u}_{m} = - i m \Omega \widehat{\delta u}_{m} - (\kappa /\etaout R_{0}^{3})
  \Gamma_{m} E_{m}(\sigma) \widehat{\delta u}_{m} + \widehat{\zeta}_{m}
\end{equation}
We can solve for $\widehat{\delta u}_{m}(\omega)$ and obtain the correlations
\begin{equation}
  \langle \widehat{\delta u}_{m}(\omega) \widehat{\delta u}_{-m}(-\omega) \rangle =
  \frac{2 k_{B}T \Gamma_{m}/\etaout
    R_{0}^{2}}{(\omega+m\Omega)^{2}+\left[\frac{\kappa \Gamma_{m}E_{m}}{\etaout R_{0}^{3}}\right]^{2}}.
\end{equation}
We have left the $\sigma$-dependence of $E_{m}$ implicit for clarity. The time
correlation function becomes ($\Delta t >0$)
\begin{equation}
  \label{eq:timecorr}
\begin{split}
  \langle \delta u_{m}(0) \delta u_{-m}(\Delta t)\rangle &= \int \frac{d\omega}{2\pi} \exp(i \omega
  \Delta t) \langle \widehat{\delta u}_{m}(\omega) \widehat{\delta u}_{-m}(-\omega)
  \rangle\\
  &= \frac{k_{B}T  R_{0}}{\kappa E_{m}}\\
  &\quad \times \exp\left[-\left({\textstyle{\frac{\kappa\Gamma_{m}E_{m}}{\etaout R_{0}^{3}}}}
  +im\Omega\right)\Delta t\right].
\end{split}
\end{equation}
with the stationary equal-time correlations
\begin{equation}
  \label{eq:statcorr}
  \langle \delta u_{m}(t)\delta u_{-m}(t)\rangle = \frac{k_{B}T  R_{0}}{\kappa E_{m}(\sigma)}.
\end{equation}
The amplitudes $u_{m}$ with different $m$ are uncorrelated
at all times. Comparison with simulation data is easier using the real Fourier
coefficients \eqref{eq:12}. The corresponding correlation functions read
\begin{equation}
  \label{eq:35}
  \begin{split}
    \langle \delta a_{m}(0) \delta a_{m}(t) \rangle &= \langle b_{m}(0)
    b_{m}(t) \rangle\\
    & = \frac{k_{B}T R_{0}}{\pi \kappa E_{m}} \exp\left(-{\textstyle{\frac{\kappa\Gamma_{m}
    E_{m}}{\eta_{\text{out}}R_{0}^{3}}}} t\right) \cos(m \dot{\gamma} t/2),\\
    \langle \delta a_{m}(0) \delta b_{m}(t) \rangle &= - \langle b_{m}(0)
    a_{m}(t) \rangle \\
    &=    \frac{k_{B}T R_{0}}{\pi \kappa E_{m}} \exp\left(-{\textstyle{\frac{\kappa\Gamma_{m}
    E_{m}}{\eta_{\text{out}}R_{0}^{3}}}} t\right) \sin(m \dot{\gamma} t/2),
  \end{split}
\end{equation}
and
\begin{equation}
  \label{eq:realstatcorr}
  \begin{split}
    \langle \delta a_{m}(0) \delta a_{m}(0) \rangle &= \langle \delta b_{m}(0)
    \delta b_{m}(0) \rangle = \frac{k_{B}TR_{0}}{\pi \kappa E_{m}(\sigma)}\\
    \langle \delta a_{m}(0) \delta b_{m}(0) \rangle &= 0
  \end{split}
\end{equation}
The fluctuating $u_{m}$ contribute to the excess length according to
Eq.~(\ref{eq:length}). Although the length constraint cannot be obeyed
exactly with a constant tension, we determine $\sigma$ such that the
constraint (\ref{eq:length}) is fulfilled on average. The total excess length
has a systematic and a fluctuating part 
\begin{equation}
  \label{eq:36}
  \Delta = \bar{\Delta}(\sigma) + \sum_{m \ge 2} \Delta_{m}(\sigma),
\end{equation}
with
\begin{equation}
  \label{eq:37}
  \begin{split}
    \bar{\Delta}(\sigma) &\equiv \sum_{m > 0} (m^{2}-1) \frac{D_{m}^{2}
      |\Phi^{\infty}_{m}|^{2}}{\Gamma_{m}^{2}E_{m}(\sigma)^{2}+
      m^{2}\tilde{\Omega}^{2}} \\
    &= \frac{3\pi}{2}\frac{\chi^{2}}{(5/2+\sigma)^{2}+\chi^{2}(1+\lambda)^{2}}.
  \end{split}
\end{equation}
and 
\begin{equation}
  \label{eq:38}
    \Delta_{m}(\sigma) \equiv  (m^{2}-1)\langle |\delta u_{m}|^{2}\rangle =  \frac{k_{B}T R_{0}}{\kappa (m^{2}-3/2+\sigma)}
\end{equation}
Thus $\sigma$ is determined implicitly by the solution of Eq.~(\ref{eq:36}). For
future reference, we note that the contribution of the fluctuating parts to the
excess length can be determined analytically to be
\begin{equation}
  \begin{split}
    \sum_{m \ge 2}\Delta_{m}(\sigma) &= \frac{k_{B}T R_{0}}{\kappa\sqrt{2} \left(4 \sigma ^2-8 \sigma
        +3\right)}
    \left[\sqrt{2} (7-6 \sigma )\right.\\
    &\quad \left. +\pi \sqrt{3-2\sigma } (2\sigma -1) \cot
      \left(\pi \sqrt{3/2-\sigma }\right)\right].
  \end{split}
\end{equation}
While this expression is exact, its behavior as a function of $\sigma$ is not
obvious (for example, the ``singularities'' at $\sigma=1/2$ and $\sigma =
3/2$ are only apparent). We therefore give the leading asymptotic behavior 
\begin{equation}
  \label{eq:39}
  \sum_{m\ge 2} \Delta_{m}(\sigma) \approx \frac{k_{B}T R_{0}}{\kappa} 
    \begin{cases}
      1/(\sigma+5/2) + 25/48& \sigma
      \rightarrow -5/2\\
      \pi (4 \sigma)^{-1/2} & \sigma \rightarrow \infty.
    \end{cases}
\end{equation}

\subsection{Zero temperature}
\label{sec:zero-temperature}

At large shear rates, nearly the entire excess length is stored in the
systematic part $\bar{\Delta}$. As a crossover shear rate $\chi_{c}$, we can
define the shear rate at which the two contributions in condition
\eqref{eq:36} become equal 
\begin{equation}
\label{eq:40}  \bar{\Delta}(\sigma,\chi_{c}) \equiv \sum_{m\ge 2} \Delta_{m}(\sigma) = \frac{\Delta}{2}. 
\end{equation}
This set of equations must be solved numerically for each $\Delta$. In the
limit $\chi \gg \chi_{c}$ we can ignore the thermal forces. In this case, the
equation of motion (\ref{eq:langevin}) becomes the deterministic Eq.~\eqref{eq:eomdet}, and the tension is
determined by Eq.~(\ref{eq:tenscons}) with $\zeta_{m}= 0$.

We can easily obtain the stationary state from $\partial_{t} u^{0}_{m} = 0$, i.e.
\begin{equation}
\label{eq:ustat}
  u^{0}_{m} = \frac{\etaout R_{0}^{3}}{\kappa} \frac{D_{m}}{\Gamma_{m}E_{m}(\sigma_{0}) + i m
    \tilde{\Omega}} \Phi^{\infty}_{m}.
\end{equation}
The homogeneous tension $\sigma_{0}$  is
determined from the length constraint \eqref{eq:length}. In the case of
constant linear shear
flow, only the $m=\pm 2$ components are non-zero and are equal in magnitude. The length constraint thus
reads $|u_{\pm2}| = (\Delta/3)^{1/2}$, or
\begin{equation}
  \begin{split}
    \Delta &= 3\frac{D_{2}^{2}}{\Gamma_{2}^{2}E_{2}^{2}+4\Omega^{2}} \frac{2\pi
      \chi^{2}}{64} \\
    &= \frac{3\pi}{2(1+\lambda)^{2}}
    \frac{\chi^{2}}{(5/2+\sigma_{0})^{2}/(9(1+\lambda))^{2}+\chi^{2}}.
  \end{split}
\end{equation}
The homogeneous tension in the
stationary state is thus given by
\begin{equation}
  \label{eq:41}
  \sigma_{0} = -5/2+ 3 \chi (1+\lambda) \left[\frac{3\pi}{2\Delta(1+\lambda)^{2}}-1\right]^{1/2}.
\end{equation}
$E_{2}$ vanishes at a critical viscosity ratio 
\begin{equation}
  \label{eq:42}
  \lambda_{c} = \sqrt{\frac{3 \pi}{2 \Delta}}-1. 
\end{equation}
This corresponds to a tank-treading to tumbling transition, as can be seen
when we allow for time-dependent $\sigma_{0}$: 
In linear shear flow, only
the $m=2$ modes are excited. In the long time
limit we can therefore assume that all other modes have decayed. In analogy
with the 3d treatment \cite{misbah2006}, we can write $u_{2}$ in polar form
\begin{equation}
  u_{2} \equiv (\Delta/3)^{1/2} \exp(-2 i \Theta),\label{eq:54}
\end{equation}
where $\Theta$ is the inclination angle of the
vesicle with respect to the shear direction. Taking the real and imaginary part of
Eq.~(\ref{eq:eomdet}) gives the familiar Jeffery's equation \cite{jeffery1922}
\begin{equation}
  \label{eq:43}
  \dot{\Theta} =  \dot{\gamma} \left[- \frac{1}{2} +
    \frac{1}{2} \frac{\sqrt{3\pi}}{(\lambda+1)\sqrt{2\Delta}} \cos(2\Theta)\right].
\end{equation}
For  $\lambda < \lambda_{c}$, Eq.~(\ref{eq:43}) admits two stationary
solutions, of which only the positive  is linearly stable 
\begin{equation}
  \label{eq:44}
  \Theta_{0} \equiv \frac{1}{2} \arccos\left(\frac{(\lambda+1)^{2}2\Delta}{3\pi}\right)^{1/2}.
\end{equation}
This corresponds to stationary tank-treading motion, where the tank-treading
frequency at zeroth order is given by the external flow
\begin{equation}
  \label{eq:45}
  \omega_{\text{tt}}^{\text{QC}} \equiv \frac{\dot{\gamma}}{2}.
\end{equation}
For $\lambda>\lambda_{c}$, the right hand side of Eq.~(\ref{eq:43}) is always
negative, and the vesicle starts to tumble. In two dimensions, no analogy to a
swinging motion (cf. Refs.
\cite{misbah2006,kantsler2006a,noguchi2007,lebedev2007}) exists, since the
volume and length constraint already uniquely determine the shape of an
ellipse.

\subsection{First-order correction to the large shear-rate limit}
\label{sec:first-order-corr}
In the mean-field approach the tension $\sigma$ is assumed constant and all
modes fluctuate independently with amplitudes given by
Eq.~(\ref{eq:statcorr}). In this picture, the length constraint is not fulfilled
rigorously but only on average. For strictly enforced length constraint the
tension must fluctuate according to Eq.~(\ref{eq:tenscons}), which
induces correlations between the deformation amplitudes. While this general effect is
worth considering in its own right, here we concentrate on the much simpler
large shear rate (or low temperature) limit as a perturbation of the
deterministic solution. 

At $T=0$, the whole excess length $\Delta$ is stored in the $|m|=2$ mode.
Perturbing the modulus of the amplitude $|u_{2}|$ alters the excess length
$\Delta$ to first order and is prohibited by the constraint \eqref{eq:2}.
Perturbing the other modes alters $\Delta$ only to second order. At low
temperature, we can therefore assume the polar decomposition~\eqref{eq:54}.
Taking the real and imaginary part of the equation of motion
\eqref{eq:langevin}, we arrive at a Langevin equation for the inclination
angle
\begin{equation}
  \label{eq:55}
  \dot{\Theta} =  \dot{\gamma} \left[- \frac{1}{2} +
    \frac{1}{2} \frac{\sqrt{3\pi}}{(\lambda+1)\sqrt{2\Delta}}
    \cos(2\Theta)\right] + \xi,
\end{equation}
where the noise term
\begin{equation}
  \label{eq:56}
  \xi \equiv \sqrt{\frac{3}{4\Delta}} \Imaginary[\zeta_{2} \exp(2i\Theta)]
\end{equation}
is Gaussian and delta-correlated
\begin{equation}
  \label{eq:57}
  \langle \xi(t) \xi(0) \rangle = \frac{3}{4\Delta} \Gamma_{2} k_{B} T \delta(t).
\end{equation}
In the stationary regime $\Theta$ fluctuates around the mean value
\begin{equation}
  \label{eq:58}
  \Theta \equiv \Theta_{0} + \Delta \Theta,
\end{equation}
where $\Theta_{0}$ is given by Eq.~\eqref{eq:44}. For small $\Delta \Theta$ we
can expand Eq.~\eqref{eq:55} to obtain
\begin{equation}
  \label{eq:59}
  \dot{\Delta \Theta} = - \dot{\gamma}\sqrt{\frac{2\pi}{3\Delta}} \frac{1}{\lambda+1}
  \sin(2\Theta_{0}) \Delta \Theta + \xi.
\end{equation}
This implies the stationary correlations
\begin{equation}
  \label{eq:60}
  \langle \Delta \Theta^{2} \rangle = \frac{R_{0} k_{B}T}{\kappa \chi
    \Delta^{1/2}}
  \frac{1}{8}\left[\frac{3\pi}{2}-(\lambda+1)^{2}\Delta\right]^{-1/2} =
  \frac{3 k_{B}T R_{0}}{8 \kappa \Delta E_{2}(\sigma_{0})}, 
\end{equation}
where we have used Eq.~\eqref{eq:44}. For small $\Delta$ we read off
\begin{equation}
  \label{eq:62}
  \langle \Delta \Theta^{2} \rangle^{1/2} \approx
  \left(\frac{1}{129\pi}\right)^{1/4} \left(\frac{\kappa \chi
      \Delta^{1/2}}{R_{0}k_{B}T}\right)^{-1/2}\approx0.20 \left(\frac{\kappa \chi
      \Delta^{1/2}}{R_{0}k_{B}T}\right)^{-1/2}.
\end{equation}
Finally, we calculate the fluctuations of the Fourier modes $a_{2}, b_{2}$.
The polar expansion \eqref{eq:54} implies
\begin{equation}
  \label{eq:61}
  \begin{split}
    a_{2} &= \sqrt{\frac{2\Delta}{3\pi}} \cos(2\Theta),\\
    b_{2} &= \sqrt{\frac{2\Delta}{3\pi}} \sin(2\Theta).
  \end{split}
\end{equation}
We derive the correlation functions of the $m=2$ modes from Eq.~\eqref{eq:60}
to be
\begin{equation}
 \label{eq:52}
  \begin{split}
    \langle \delta a_{2} \delta a_{2}\rangle &=\frac{k_{B}T R_{0}}{\pi \kappa
      E_{2}(\sigma_{0})} \cos^{2}(2\Theta_{0}),\\
    \langle \delta a_{2} \delta b_{2}\rangle &=\frac{k_{B}T R_{0}}{\pi \kappa E_{2}(\sigma_{0})} \cos(2\Theta_{0})\sin(2\Theta_{0}),\\
    \langle \delta b_{2} \delta b_{2}\rangle &=\frac{k_{B}T R_{0}}{\pi \kappa
      E_{2}(\sigma_{0})} \sin^{2}(2\Theta_{0}).
  \end{split}
\end{equation}

\section{Keller-Skalak theory}
\label{sec:keller-skalak-theory}

In the theory of Keller and Skalak \cite{keller1982}, a three-dimensional vesicle
is assumed to have a fixed ellipsoidal shape
\begin{equation}
(x_1/a_1)^2+(x_2/a_2)^2+(x_3/a_3)^2=1,
\end{equation}
where the $a_i$ are the semi-axes of the ellipsoid, and the 
coordinate axes $x_i$ point along its principal directions.
The $x_1$ and $x_2$ axes, with $a_1>a_2$, are chosen to lie in the $xy$ plane
and are rotated through an angle $\Theta$ with respect to the $x$
and $y$ axes.
The components of the undisturbed shear flow are $(\dot{\gamma} y,0,0)$.
The velocity field at the membrane is assumed to be
\begin{equation}
\Vv =\omega^{\text{KS}}_{\text{tt}} \left( -(a_1/a_2)x_2,(a_2/a_1)x_1,0 \right),
\end{equation}
where $\omega^{\text{KS}}_{\text{tt}}$ is a parameter having the dimensions of a frequency.
The energy supplied by the external flow has to be balanced with the
energy dissipated inside the vesicle.
The motion of the vesicle derived from this energy balance reads \cite{keller1982}
\begin{equation}
\frac{d \Theta}{d t} = -\frac{\dot{\gamma}}{2} + B \cos(2 \Theta),
\label{ks1}
\end{equation}
with
\begin{equation}
B = \frac{\dot{\gamma}}{1+r_2^2} 
\Big \{ 
\frac{(1-r_2^2)^2 [z_2(1-\lambda)-2]-8r_2^2}{2(1-r_2^2)[z_2(1-\lambda)-2]}
\Big \}
\label{ks2}
\end{equation}
and
\begin{equation} 
\omega^{\text{KS}}_{\text{tt}} = 2 \dot{\gamma} \frac{r_2(1+r_2^2)}{(1-r_2^2)^2 [z_2(1-\lambda)-2]-8r_2^2}.
\label{ks3}
\end{equation}
The factors appearing in Eqs.~(\ref{ks1})-(\ref{ks3}) are given by
\begin{equation}
  \begin{array}{lll}
r_2 \equiv a_2/a_1,& r_{3} \equiv a_{3} / a_{1}, &z_2 \equiv g_3^{'}(\alpha_1^2+\alpha_2^2),\\
\alpha_1\equiv r_2^{-1/3} r_3^{-1/3},&
\alpha_2\equiv r_2^{2/3} r_3^{-1/3},&
\alpha_3\equiv r_2^{-1/3} r_3^{2/3},
\end{array}
\end{equation}
and
\begin{equation}
g_3^{'}\equiv \int_0^{\infty} (\alpha_1^2+s)^{-3/2} (\alpha_2^2+s)^{-3/2} 
(\alpha_3^2+s)^{-1/2}ds.
\end{equation}
For $B>\dot{\gamma}/2$, we obtain a steady tank-treading angle
\begin{equation}
  \label{eq:46}
  \Theta = \frac{1}{2} \arccos\left(\frac{\dot{\gamma}}{2B}\right).
\end{equation}
We calculate the inclination angle $\Theta$ and
the tank-trading frequency $\omega^{\text{KS}}_{\text{tt}}$ by adapting the
Keller-Skalak theory to two dimensions. We numerically solve
Eqs.~(\ref{ks1})-(\ref{ks3}) in the limit $r_3 \rightarrow +\infty$ keeping
$r_2$ finite, which formally corresponds to an ellipsoid with an infinite
semi-axis in the $z$ direction.

\section{Simulation method}
\label{sec:simulation-method}
A 2d vesicle model system was simulated using the multi-particle collision (MPC)
dynamics \cite{malevanets1999,noguchi2004,yeomans2006}.  
In this method the fluid is not treated on a continuum level, but rather by a
stochastic dynamics of effective fluid particles. 

\subsection{Solvent dynamics}
We consider a two-dimensional system made of
$N_s$ identical particles of mass $m_s$ whose positions ${\bf r}_i(t)$ and
velocities ${\bf v}_i(t)$, $i=1,2,\ldots,N_s$, are continuous variables.
The time is discretized in intervals $\Delta t_s$. 
The evolution occurs in two
consecutive steps, streaming and collision.
In the streaming step, particles move ballistically,
\begin{equation}
{\bf r}_i(t+\Delta t_s)={\bf r}_i(t)+{\bf v}_i(t) \Delta t_s.
\label{eq.prop}
\end{equation}
For the collision step, the system is divided into the cells of a regular square
lattice of mesh size $a$. Each of these cells is the interaction area where an
instantaneous multi-particle collision occurs, which changes particles
velocities as \cite{malevanets1999}
\begin{equation}
{\bf v}_i(t+\Delta t_s)={\bf u}(t)+ 
\mathsf{\Omega}[{\bf v}_i(t)-{\bf u}(t)],
\label{eq.coll}
\end{equation}
where ${\bf u}$ is the average velocity of
the colliding particles in a cell.
The velocity field ${\bf u}$ is considered to be
the macroscopic velocity of the fluid and it is assumed to have the
coordinates of the center of the cell. $\mathsf{\Omega}$ denotes a
stochastic rotation matrix which rotates,
with equal probability, by an angle of either $+\alpha$ or $-\alpha$. 
The collisions are
performed simultaneously on all the particles in a cell with the same rotation
$\mathsf{\Omega}$, but $\mathsf{\Omega}$
may differ from cell to cell. The local momentum and kinetic
energy are conserved under this dynamics. The kinetic energy of particles fixes
the temperature $k_B T$, where $k_B$ is the Boltzmann constant, via the 
equipartition theorem.

It was shown in Ref.~\cite{ripoll2004} that a proper description of
hydrodynamics in MPC requires large Schimdt numbers. This can be
accomplished by choosing a mean-free path 
$l=\Delta t_s \sqrt{k_B T /m_s}$, which is small compared to the cell size $a$.
It is known that a value of $l$ much smaller than $a$
breaks the Galilean invariance \cite{ihle2001} and
that this problem can be solved by applying a random shift procedure \cite{ihle2001}. 
The viscosity of the solvent fluid is \cite{kikuchi2003,ihle2004}
\begin{equation}
  \begin{split}
    \eta&=\left [ \frac{l}{2 a} \Big [ \frac{n_c^2}{(n_c-1+e^{-n_c})
        \sin^2\alpha} - n_c \Big ]\right. \\
    &\quad \left. + \frac{a}{12 l} (n_c -1+e^{-n_c}) (1 -
      \cos \alpha) \right ] \frac{\sqrt{m_s k_B T}}{a},
  \end{split}
\label{visc}
\end{equation}
with particle density $\rho=n_c m_s / a^2$ and number $n_c$ of particles 
per cell.

In order to enforce shear flow, we place our system of size $L_x \times L_y$
between two horizontal walls. The upper and the lower walls slide along the
$x$ direction with velocities ${\bf v}_{wall} = (v_{wall}, 0)$ and $-{\bf
  v}_{wall}$, respectively, with $v_{wall} > 0$. Periodic boundary conditions
are used along the $x$ direction. Along the $y$ direction, we use a modified
bounce-back boundary condition which consists in requiring that particles
hitting the walls change their velocities according to ${\bf
  v}_i \rightarrow 2 {\bf v}_{wall} - {\bf v}_i$.  Together
with virtual particles in partly filled cells at
walls, this describes no-slip boundary conditions very well \cite{lamura2001a,lamura2002}. A linear
flow profile $(u_x, u_y)=(\dot{\gamma} y, 0)$ is obtained with shear rate
$\dot{\gamma} = 2 v_{wall} / L_y$, with the walls placed at $y=\pm L_y/2$. The
relative velocities in the collision cells are rescaled after each time step
$\Delta t_s$ in order to keep the temperature constant in the (driven) system.

\subsection{Membrane model}

The vesicle membrane is modeled by connecting $N_{p}$ beads of mass $m_{p}$
successively with bonds into a closed ring. Neighboring beads along the closed
chain are connected to each other with the harmonic potential
\begin{equation}
  U_{\text{bond}} \equiv \frac{k_{h}}{2}\sum_{i=1}^{N_{p}}
  \frac{(|\Vr_{i}-\Vr_{i-1}|-r_{0})^{2}}{r_{0}^{2}},
\end{equation}
where $k_{h}$ is a spring constant, $\Vr_{i}$ is the position vector of the
$i$-th bead, and $r_{0}$ is the average bond length. The bending energy
\eqref{eq:15} is modeled on the discrete level by a bending potential
\begin{equation}
  \label{eq:53}
  U_{\text{bend}}\equiv \frac{\kappa}{r_{0}}\sum_{i=1}^{N_{p}}(1-\cos \beta_{i}),
\end{equation}
where $\beta_{i}$ is the angle between successive bonds. The fluid modeled
with the MPC method is compressible. To enforce the area constraint in the
presence of thermal and hydrodynamic forces, we add a constraint potential
\begin{equation}
  U_{\text{area}} \equiv \frac{k_{A}}{2} \frac{(A-A_{0})^{2}}{r_{0}^{4}}.\label{area_pot}
\end{equation}


\subsection{Coupling of membrane and solvent dynamics}

The membrane-solvent interaction must prevent solvent particles from crossing
the membrane and enforce no-slip boundary conditions on the membrane.
Therefore we place hard disks centered on the membrane beads. The disk radius
$r_p$ is set in order to ensure overlapping of disks and a complete coverage
of the membrane. The exchange of momentum between the solvent particles and
the membrane occurs in the following way. After updating beads positions and
velocities via molecular dynamics (MD), we freely stream all the solvent
particles. We then execute bounce-back scattering between solvent and membrane
disks only when a solvent particle $j$ and a disk $i$ satisfy the conditions
$|{\bf r}_i-{\bf r}_j| < r_p$ and $ ({\bf r}_i-{\bf r}_j) \cdot ({\bf
  v}_i-{\bf v}_j) < 0$. This means that if the two collision partners $i$ and
$j$ overlap and move towards each other, then their velocities are updated
according to
\begin{eqnarray}
{\bf v}_i &\rightarrow& 
{\bf v}_i - 2 \frac{m_s}{m_s+m_p} ({\bf v}_i-{\bf v}_j), \nonumber \\ 
{\bf v}_j &\rightarrow& 
{\bf v}_j + 2 \frac{m_p}{m_s+m_p} ({\bf v}_i-{\bf v}_j).
\end{eqnarray} 
To avoid that a solvent particle moves too far inside a disk, we require that $l
\ll r_p$. The collision step (\ref{eq.coll}) is performed only on those
solvent particles which did not scatter. If the collision step were executed
also on the scattered solvent particles, they might continue to collide with the same disk in the
next time step. The fluids in the interior and exterior of the vesicle are
taken to be the same, in particular to have the same viscosity.

A chain of disks of finite radius $r_{p}$ has an inner length available to the
solvent particles which is smaller than the outer length. Since the solvent
has the same density inside and outside, the outer fluid exerts a compression
force on the membrane until the inner density increases so that an expansion
force compensates the compression one. It is straightforward to show
\cite{noguchi2005a} that the density increase is $\Delta \rho / \rho = 2 r_p /
R^{*}$ where $2 r_p$ is the effective membrane thickness. This requires that
$R_0$ is large enough compared to the disk radius $r_p$ to reduce such
compression effects. The number of solvent particles placed inside the vesicle
fixes an average area. However, since the MPC fluid is compressible, shear and
bending rigidity effects may change the area $A$. For this reason the
constraint potential (\ref{area_pot}) is introduced to keep the area constant.

\subsection{Parameters}

In experiments with vesicles in shear flow, inertial effects are negligible
since the Reynolds number $Re \equiv \dot{\gamma} \rho {R^{*}}^2 / \etaout$ is very small.
We express our results using the reduced area $A^*=A_0/\pi {R^{*}}^2$, defined
in Eq.~(\ref{eq:4}), and the reduced shear rate $\chi = \dot{\gamma}\etaout
R_{0}^{3}/\kappa$, see Eq.~(\ref{eq:22}), as relevant dimensionless quantities.

\begin{figure}
  \centering
  \includegraphics[angle=-90,width=0.85\linewidth]{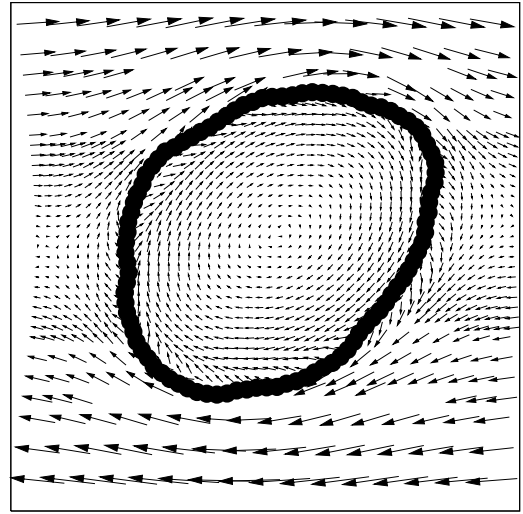}
  \caption{Snapshot of the vesicle and the velocity field around it, taken
    from simulation data for reduced area $A^{*} = 0.95$ and reduced shear
    rate $\chi \simeq 5.6$ (see Eq.~\eqref{eq:22}). The disks represent the
    beads forming the membrane and are plotted to scale.}
  \label{fig:velocity}
\end{figure}

We set $\alpha=\pi/4$, $n_c=10$, and $l =0.008 a$. This implies a viscosity
$\etaout = \etain \simeq 28.0 \sqrt{m_{s} k_{B}T} / a$. We use $L_x=150 a$, $L_y=90
a$, $R^{*}=15.3 a$, and $v_{wall}$ such that $Re < 0.1$ for all the cases we
considered with $0.5 \le \chi \le 10.0$. Finally, we set $m_p=10 m_s$,
$N_p=96$, $\Delta t_p = \Delta t_s / 20$, $r_p=0.9 a$, $r_0=a$, $\kappa=40 k_B
T a$, $k_{A}=0.5 k_B T$, $k_h=4000 k_B T$. The area $A_0$ is chosen in such a way that $0.7 \leq
A^* \leq 0.95$. With the choices for $k_A$ and $k_h$ the area and the length
of the vesicle are kept constant with a deviation of less than $1\%$ of the
target values for all simulated systems. A snapshot of a simulated vesicle
and the resulting velocity field for the reduced area $A^{*}=0.95$ and reduced
shear rate $\chi = 5.6$ is shown in Fig.~\ref{fig:velocity}. 

\section{Results and discussion}
\label{sec:comp-with-simul}

\subsection{Stationary deformations}
\label{sec:stat-deform}
\begin{figure}
  \centering
  \includegraphics[angle=-90,width=\linewidth]{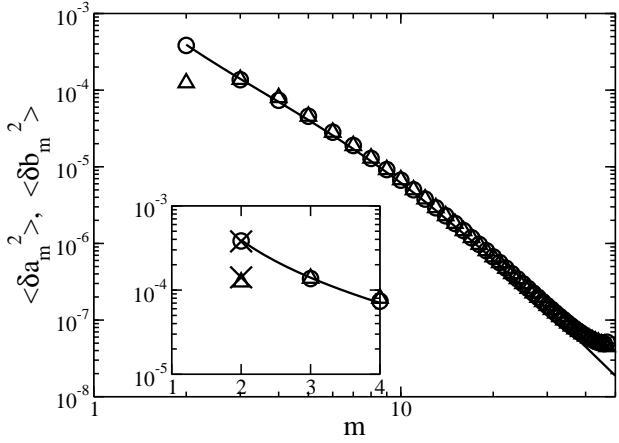}
  \caption{Stationary correlations $\langle \delta a_{m}^{2} \rangle$
    (triangles) and $\langle \delta b_{m}^{2} \rangle$ (circles) as a function of
    mode number $m$ on a double logarithmic scale. The solid line corresponds
    to a fit of Eq.~\eqref{eq:realstatcorr} with
    $\Sigma_{\text{fit}}\simeq 103\kappa/R_{0}^{2}$. The inset highlights $\langle \delta
    a_{2}^{2}\rangle$, $\langle \delta b_{2}^{2} \rangle$ in comparison with
    the low-temperature expansion Eq.~\eqref{eq:52} (crosses).
    Simulation parameters are $\Delta = 0.163$ (corresponding to $A^{*}=0.95$)
    and $\chi \simeq 9.3$.}
  \label{fig:correlations}
\end{figure}

In Fig.~\ref{fig:correlations} we show the stationary deformation
correlations $\langle \delta a_{m}^{2}\rangle$, $\langle \delta
b_{m}^{2}\rangle$ as a function of the mode number $m$ for $\Delta = 0.163$ and
$\chi\simeq 9.3$. We also show a fit of these correlations for $m\ge 3$ 
with the
theoretical prediction \eqref{eq:realstatcorr}. From the fit we can extract
the tension $\Sigma$. In this particular example we obtain
$\Sigma_{\text{fit}} \simeq 103 \kappa/R_{0}^{2}$, whereas theory predicts
$\Sigma_{\text{theor}} \simeq 113 \kappa/R_{0}^{2}$ from
Eqs.~\eqref{eq:36}--\eqref{eq:38}. 

The mean field treatment \eqref{eq:realstatcorr} predicts $\langle \delta
a_{m}^{2}\rangle = \langle \delta b_{m}^{2}\rangle$ for all $m$. We can see
that this holds only for $m \ge 3$. As explained in
Sec.~\ref{sec:first-order-corr}, this is due to the fluctuations in the line
tension. In the inset
of Fig.~\ref{fig:correlations}, we compare $\langle \delta a_{2}^{2}
\rangle$, $\langle \delta b_{2}^{2}\rangle$ with the low temperature expansion
(\ref{eq:52}), with very good agreement. 

\subsection{Tension vs shear rate}
\label{sec:tension-vs-shear}

\begin{figure}
  \centering
  \includegraphics[angle=-90,width=\linewidth]{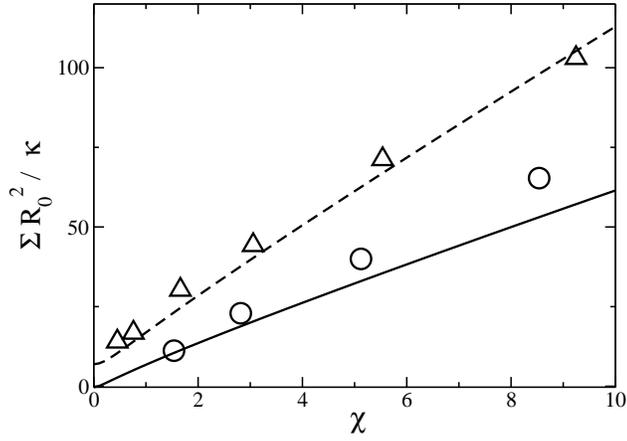}
  \caption{Dimensionless tension $\Sigma R_{0}^{2}/\kappa$ as a function of
    reduced shear rate
    $\chi$. Symbols denote the fitted tensions $\sigma_{\text{fit}}$ extracted
    from fluctuation spectra for excess lengths $\Delta=0.163$ (triangles) and
    $\Delta = 0.340$ (circles), corresponding to $A^{*}=0.95$ and $A^{*}=0.9$, respectively (compare Fig.~\ref{fig:correlations}). The solid and
    dashed lines show the corresponding numerical solution of
    Eqs.~\eqref{eq:36}--\eqref{eq:38}.} 
  \label{fig:chisigma}
\end{figure}

Fig.~\ref{fig:chisigma} shows the extracted dimensionless tensions
$\Sigma_{\text{fit}} R_{0}^{2}/\kappa$
for different dimensionless shear rates $0 \le \chi \le 10$ and for two
different excess lengths $\Delta = 0.163$ and $\Delta = 0.340$. The agreement
with the theoretical prediction from a numerical solution of
Eqs.~\eqref{eq:36}--\eqref{eq:38} is satisfactory. The fact that this
function is nearly a straight line implies that the large-shear-rate
approximation (\ref{eq:7}) is valid down to small shear rates. We find a crossover shear rate $\chi_{c}$, below which
there are deviations from a linear behavior. Theoretically, Eq.~\eqref{eq:40}
gives an order-of-magnitude estimate of $\chi_{c}\simeq 3.09$ for $\Delta =
0.163$ and $\chi_{c} \simeq 0.99$ for $\Delta = 0.340$.

\subsection{Autocorrelation function}
\label{sec:autoc-funct}

\begin{figure}
  \centering
  \includegraphics[angle=-90,width=\linewidth]{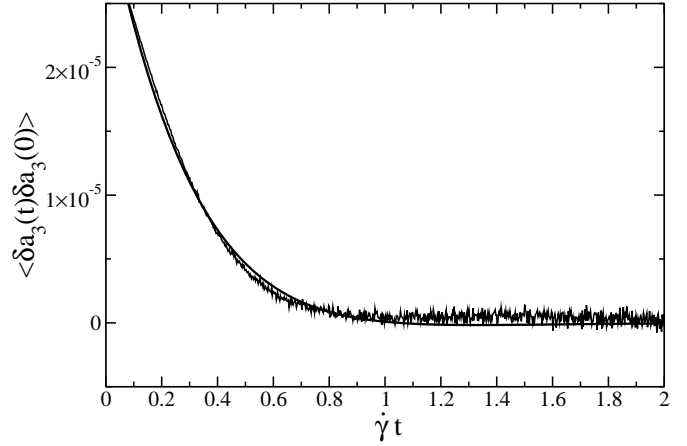}
  \caption{Autocorrelation function $\langle \delta a_{3}(t) \delta
    a_{3}(0)\rangle$ of the $a_3$ mode as a function of dimensionless time
    $\dot{\gamma}t$. Simulation parameters are $\Delta = 0.163, \chi = 5.6$.
    The solid line shows a fit of Eq.~\eqref{eq:35}.}
  \label{fig:autocorr}
\end{figure}

In Fig.~\ref{fig:autocorr}, the time autocorrelation function $\langle
a_{3}(t) a_{3}(0)\rangle$ is shown as a function of dimensionless time
$\dot{\gamma} t$. The data follows the expected exponential decay
(\ref{eq:35}) very well. From the amplitude $\langle \delta a_{3}(0) \delta
a_{3}(0) \rangle$ we can extract a tension $\Sigma \simeq 65 \kappa
/R_{0}^{2}$, while from the time constant we deduce $\Sigma \simeq 44
\kappa/R_{0}^{2}$. Theory predicts $\Sigma_{\text{theor}}\simeq
73\kappa/R_{0}^{2}$. Given the rather noisy data, this agreement seems
reasonable. For moderate shear rates the autocorrelation function has decayed
before the oscillations implied by Eq.~(\ref{eq:35}) become noticeable. Even
for the large shear rate $\chi \simeq 5.6$ used in Fig.~\ref{fig:autocorr}, the
oscillations are barely visible. For the same reason the build up of cross
correlation $\langle a(t) b(0) \rangle$ is hidden in the numerical noise.

\subsection{Inclination angle}
\label{sec:inclination-angle}

\begin{figure}
  \centering
 \includegraphics[angle=-90,width=\linewidth]{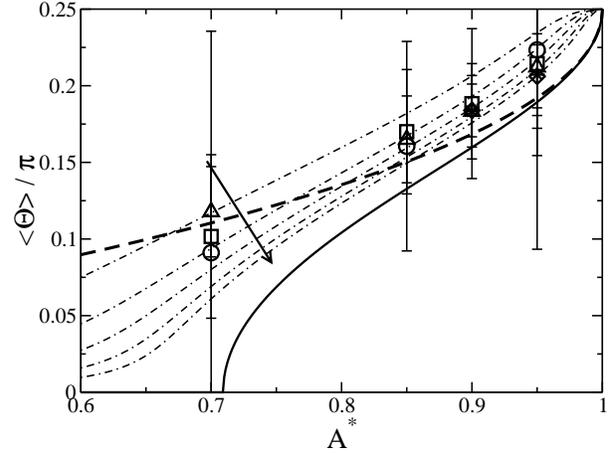}
 \caption{Scaled average inclination angle $\langle \Theta\rangle / \pi$ as a
   function of reduced area $A^*$. Symbols with error bars show simulation
   data for different values of the reduced shear rate $\chi=10.0 {A^{*}}^{3/2}$ (diamonds),
   $\chi=6.0 {A^{*}}^{3/2}$ (stars), $\chi = 3.3 {A^{*}}^{3/2}$ (triangles),
   $\chi = 1.8 {A^{*}}^{3/2}$ (squares), $\chi =
   0.8 {A^{*}}^{3/2}$ (circles). The continuous line corresponds to the deterministic limit
   Eq.~(\ref{eq:44}) and is independent of $\chi$. Dashed-dot lines follow
   from the mean-field Eq.~(\ref{eq:umean}) with $\chi {A^{*}}^{-3/2}\in
   \{0.8,1.8,3.3,6,10\}$ growing in the direction indicated by the arrow. The
   thick dashed line follows from the Keller-Skalak theory, see Eq.~\eqref{eq:46}.}
  \label{fig:incl-astar}
\end{figure}

We compare the averaged inclination angle $\langle \Theta \rangle$ for
different reduced areas $A^{*}$ with Eq.~(\ref{eq:44}) in Fig.~\ref{fig:incl-astar}, valid in the
quasi-circular limit. The agreement with simulation data is 
satisfactory, given the large error
bars. The 2d Keller-Skalak theory, which is also shown in the plot, gives
slightly better agreement.

\begin{figure}
  \centering
  \includegraphics[angle=-90,width=\linewidth]{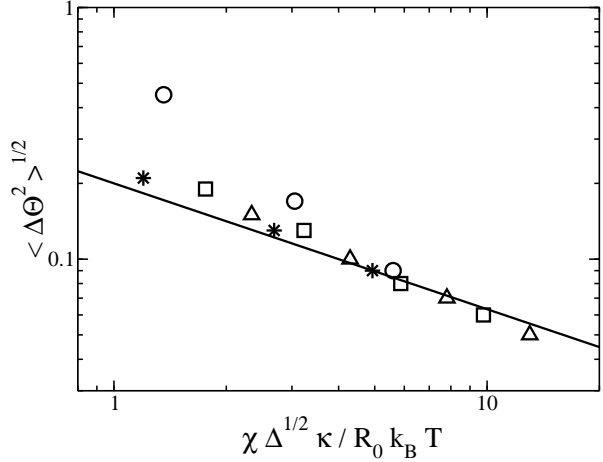}
  \caption{Fluctuations of the inclination angle $\langle \Delta
    \Theta^{1/2}\rangle^{1/2}$ as a function of reduced shear rate $\chi
    \Delta^{1/2} \kappa / (R_{0} k_{B}T)$. Symbols denote simulation data for
    different $A^{*} = 0.95$ (squares), $A^{*} = 0.90$ (triangles), $A^{*}=0.85$
    (stars), $A^{*} = 0.7$ (circles), corresponding to $\Delta \simeq 0.163$,
    $\Delta \simeq 0.34$, 
    $\Delta \simeq 0.53$, and $\Delta \simeq 1.23$, respectively. The solid
    line is the quasi-circular scaling prediction of Eq.~\eqref{eq:62}.}
  \label{fig:delta-theta-chi}
\end{figure}

In Fig.~\ref{fig:delta-theta-chi}, we show the fluctuations of the
inclination angle $\langle \Delta \Theta^{2}\rangle \equiv \langle \Theta^{2}\rangle - \langle
\Theta\rangle^{2}$ as a function of the shear rate. The theoretical scaling is
given by Eq.~(\ref{eq:62}), and is in excellent agreement for $A^{*}=0.85$
($\Delta \simeq 0.53$),
$A^{*} = 0.9$ ($\Delta \simeq 0.34$), and $A^{*} = 0.95$ ($\Delta \simeq 0.163$). The scaling of the fluctuations of $\Theta$ for $A^{*}$ differs
significantly for $A^{*}=0.7$ ($\Delta \simeq 1.23$). In the deterministic case, a vesicle with such a
low reduced area would tumble within the quasi-circular theory. This
implies that the quasi-circular approximation works well for $\Delta \le 0.53$
(corresponding to $A^{*} \ge 0.85$) in two dimensions.

\subsection{Tank-treading frequency}
\label{sec:tank-tread-freq}

\begin{figure}
  \centering
  \includegraphics[angle=-90,width=\linewidth]{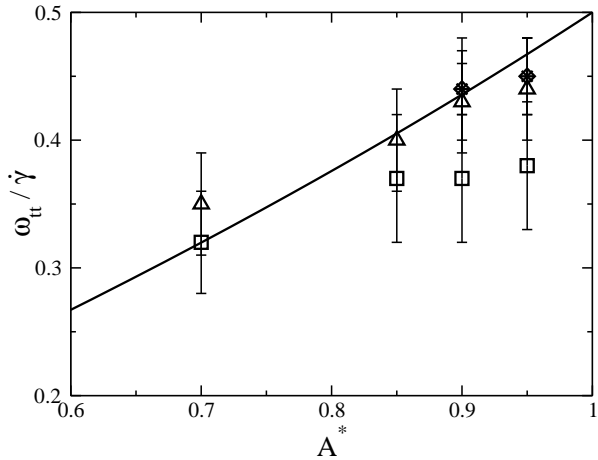}
  \caption{Tank-treading frequency $\omega_{\text{tt}}$ rescaled by
    shear rate $\dot{\gamma}$, as a function of reduced area $A^*$ for
    different values of the reduced shear rate $\chi = 10.0 {A^{*}}^{3/2}$
    (diamonds), $6.0 {A^{*}}^{3/2}$ (stars), $3.3 {A^{*}}^{3/2}$
    (triangles), $1.8 {A^{*}}^{3/2}$ (squares). The solid line follows from
    the Keller-Skalak theory, see Eq.~(\ref{ks3}).}
  \label{fig:omegatt}
\end{figure}

Finally, we show the rescaled tank-treading frequency $\omega_{\text{tt}}/\dot{\gamma}$  as a function of $A^{*}$
in Fig.~\ref{fig:omegatt}. Again, agreement with the 2d Keller-Skalak theory
is quite good, while the quasi-circular theory neglects the effect of the
vesicle shape on the flow and would predict
$\omega_{\text{tt}}^{\text{QC}}/\dot{\gamma} = 1/2$, see Eq.~\eqref{eq:45}.

\section{Summary}
\label{sec:summary}

We have studied the fluctuations and deformation of a 2d vesicle
in shear flow at finite temperature. In the limit of small deformations from a
circle, we have derived analytical Langevin-type equations of motion, which
are nonlinear due to the length constraint. A mean-field treatment allows
approximate predictions for the stationary correlation functions and time
autocorrelation functions of the deformation amplitudes, which agree
quantitatively with simulation data. Deviations of the stationary correlations from
the mean-field predictions in the lowest mode are explained quantitatively
in a low temperature expansion of the original constrained Langevin equations.
The mean inclination angle and the tank-treading frequency are better
described by a deterministic 2d Keller-Skalak theory. Fluctuations of the
inclination angle are also determined quantitatively. Theory and simulations
agree well for low excess lengths, but differ for larger excess lengths.

The good quantitative agreement of mesoscale simulations of vesicles in flow
with detailed theoretical calculations demonstrates the predictive power of
these simulation methods for more complex flow geometries. 
\begin{acknowledgement}
RF, US and GG would like to acknowledge financial support through the DFG priority program
SPP 1146 ``Micro- and Nanofluidics''.
AL and GG acknowledge fruitful discussions with H. Noguchi, M. Ripoll, 
G. Vliegenthart, and R. Winkler.
AL thanks Gerhard Gompper and co-workers for hospitality at the 
Forschungszentrum J\"{u}lich and acknowledges support from CNR through
the Short-Term Mobility Program.
\end{acknowledgement}



\bibliography{papers}

\end{document}